\documentclass[12pt]{article}
\usepackage{graphicx}
\usepackage{epsfig}
\usepackage{slashbox}
\usepackage{wrapfig}

\def\pbnr{}
\def\Authorc{M. Padmanath}
\def\speaker{Mike Peardon}
\def\Authora{Robert\ G.\ Edwards}
\def\Authorb{Nilmani\ Mathur}
\def\and{and}
\def\onbehalfof{Hadron Spectrum Collaboration}
\def\title{Excited-state spectroscopy of singly, doubly and triply-charmed baryons from lattice QCD}
\def\affiliations{School of Mathematics, Trinity College, Dublin, Ireland.}
\def\affiliationa{Jefferson Laboratory, Newport News, VA, USA.}
\def\affiliationbc{Tata Institute of Fundamental Research, Mumbai, India.}
\def\emails{mjp@maths.tcd.ie; $^*${\it \textbf{Speaker}}}
\def\emaila{edwards@jlab.org}
\def\emailb{nilmani@theory.tifr.res.in}
\def\emailc{padmanath@theory.tifr.res.in}
\newcommand\pubnumber{\pbnr}
\newcommand\pubdate{\today}
\def\Title#1{\begin{center} {\Large #1 } \end{center}}
\def\Author#1{\begin{center}{ \sc #1} \end{center}}

\newcommand{\OnBehalf}[1]{\sbox0{#1}\ifdim\wd0=0pt
        {}
	\else
	{on behalf of #1}
	\fi}
\newcommand{\SupportedBy}[1]{\sbox0{#1}\ifdim\wd0=0pt
        {}
	\else
	{\footnote{#1}}
	\fi}
\def\Speakermark{\Large $^*$ }
\def\Address#1{\begin{center}{ \it #1} \end{center}}

\newcommand\pubblock{\includegraphics[width=5cm]{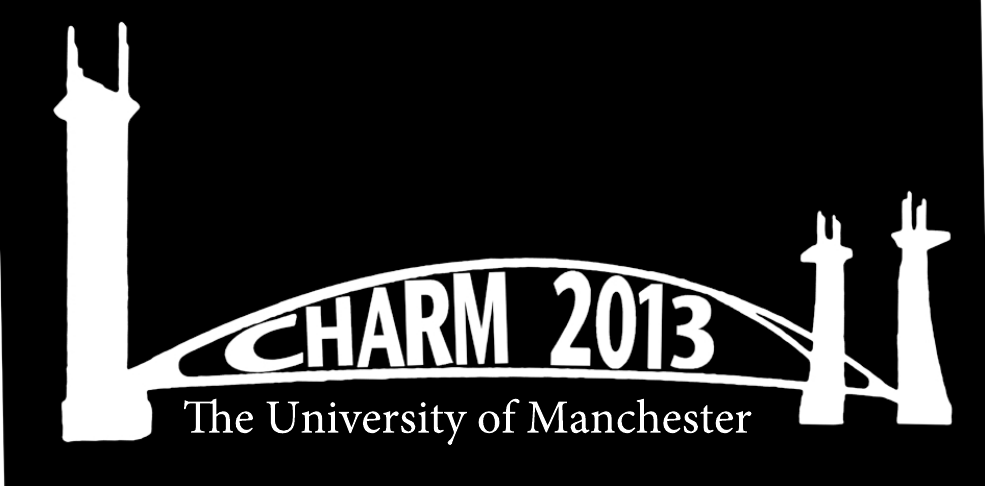}\hfill{\begin{tabular}{l} \pubnumber\\
         \pubdate  \end{tabular}}}
\newenvironment{Abstract}{\begin{quotation}  }{\end{quotation}}
\newenvironment{Presented}{\begin{quotation} \begin{center} 
             PRESENTED AT\end{center} 
      \begin{center}\begin{large}}{\end{large}\end{center} \end{quotation}}
\def\Acknowledgements{\bigskip  \bigskip \begin{center} \begin{large}
             \bf ACKNOWLEDGEMENTS \end{large}\end{center}}
\def\venue{The 6$^{th}$ International Workshop on Charm Physics\\
(CHARM 2013)\\
Manchester, UK,  31 August -- 4 September, 2013}

\newcommand\bef{\begin{figure}}
\newcommand\eef[1]{\label{fg:#1}\end{figure}}
\newcommand\fgn[1]{Figure \ref{fg:#1}}
\newcommand\bet{\begin{table}}
\newcommand\eet[1]{\label{tb:#1}\end{table}}
\newcommand\betb{\begin{center}\begin{tabular}}
\newcommand\eetb{\end{tabular}\end{center}}

\def\beq{\begin{equation}}
\def\eeq#1{\label{#1}\end{equation}}
\def\eeqn{\end{equation}}

\def\beqa{\begin{eqnarray}}
\def\eeqa#1{\label{#1}\end{eqnarray}}
\def\eeqan{\end{eqnarray}}

\let\bar=\overbar

\def\Dslash{\not{\hbox{\kern-4pt $D$}}}
\def\dslash{\not{\hbox{\kern-2pt $\del$}}}

\def\msb{{\bar{\ssstyle M \kern -1pt S}}}

\begin{document}
\begin{titlepage}
\pubblock

\vfill
\Title{\title}
\vfill
\vspace{-0.3cm}
\Author{\Authorc\SupportedBy{\emailc}}
\vspace{-0.5cm}
\Address{\affiliationbc}
\Author{\Authora\SupportedBy{\emaila},}
\vspace{-0.5cm}
\Address{\affiliationa}
\Author{\Authorb\SupportedBy{\emailb}}
\vspace{-0.5cm}
\Address{\affiliationbc}
\Address{\and}
\vspace{-0.3cm}
\Author{\speaker\SupportedBy{\emails}\Speakermark}
\vspace{-0.5cm}
\Address{\affiliations}
\Author{\OnBehalf{\onbehalfof}}
\vfill
\begin{Abstract}
We present the ground and excited state spectra of singly, doubly and
triply-charmed baryons by using dynamical lattice QCD. A large set of
baryonic operators that respect the symmetries of the lattice and are
obtained after subduction from their continuum analogues are utilized.
These operators transform as irreducible representations of SU(3)$_F$
symmetry for flavour, SU(4) symmetry for Dirac spins of quarks and
O(3) symmetry for orbital angular momenta. Using novel computational
techniques correlation functions of these operators are generated and
the variational method is exploited to extract excited states. The
lattice spectra that we obtain have baryonic states with well-defined
total spins up to 7/2 and the low lying states remarkably resemble the
expectations of quantum numbers from SU(6)$\otimes$O(3) symmetry. 
\end{Abstract}
\vfill
\begin{Presented}
\venue
\end{Presented}
\vfill
\end{titlepage}
\def\thefootnote{\fnsymbol{footnote}}
\setcounter{footnote}{0}
%

\section{Introduction}
\vspace*{-0.1in}
Heavy hadron spectroscopy finds itself in a rejuvenated phase following the discovery 
of numerous heavy hadrons in the past decade at various particle colliders, like Belle, 
BaBar, CDF, LHCb, BES-III, etc. Heavy quarkonia have been studied comprehensively 
both theoretically and experimentally. However, heavy baryons have not been explored in
great detail, though they can also provide similar information about the quark 
confinement mechanism as well as augmenting our knowledge about the nature of strong 
force by providing a clean probe of the interplay between perturbative and 
non-perturbative QCD. Experimentally only a handful of singly charmed baryons have 
been discovered, the discovery of doubly charmed baryon is controversial, whereas no 
triply heavy baryon has been observed yet \cite{PDG}. Moreover, most of the observed 
charmed baryons do not have assigned quantum numbers yet. However, it is expected that 
the large data set that will be collected in experiments at BES-III, the LHCb, and 
the planned PANDA experiment at GS/FAIR may provide significant information for 
baryons with heavy quarks. In light of these existing and future experimental prospects 
on charm baryon studies, it is desirable to have model independent predictions from 
first principles calculations, such as from lattice QCD. Results from such calculations 
will naturally provide crucial inputs to the future experimental discovery and can 
be compared with those obtained from potential models which have been very successful 
in the case of charmonia. Details of various potential model calculations for charm 
baryons can be found in Refs. \cite{Capstick:2000qj, Klempt:2009pi, Crede:2013kia}. 
Till very recently, lattice QCD results on charmed baryons included only the ground 
states with spin up to ${3\over 2}$ \cite{Briceno:2012wt,Basak:2012py,Namekawa:2013vu,Bali:2012ua}.
In this proceeding, we present our results on comprehensive spectra of doubly and triply charmed 
baryons \cite{Padmanath:2013zfa, Padmanath:2013laa}, along with the preliminary 
results on singly charmed baryon spectra.

\vspace{-0.5cm}
\section{Numerical details}
\vspace{-0.2cm}
We utilized the ensemble of dynamical anisotropic gauge-field configurations generated by 
the Hadron Spectrum Collaboration (HSC) to extract highly excited hadron spectra. Adopting 
a large anisotropy co-efficient $\xi=a_s/a_t=3.5$, with  $a_t~m_c\ll1$, we could use the 
standard relativistic formulation of fermions for all the quark flavors from light to 
charm. Along with an $O(a)$-improved gauge action, we used an anisotropic 
Shekholeslami-Wohlert action with tree-level tadpole improvement and stout-smeared spatial 
links for the $N_f=2+1$ dynamical flavours fermionic fields and the valence fermionic fields. 
The temporal lattice spacing, $a_t^{-1}=5.67$GeV, was determined by equating the $m_{\Omega}$ 
to its physical value, resulting in a lattice spatial extension of 1.9 fm, which presumably be 
sufficiently large for a study of charmed baryons. We used an ensemble of 96  
configurations with a temporal extension equal to 128. The pion masses in these lattices 
were determined to be 391 MeV. More details of the formulation of actions as well as the techniques 
used to determine the anisotropy parameters can be found in Refs. \cite{Edwards:2008ja, 
Lin:2008pr}.

\vspace{-0.5cm}
\section{Operator construction and spin identification}
\vspace{-0.2cm}
\noindent We use a large basis of operators, constructed employing the derivative-based 
operator construction formalism \cite{Edwards:2011jj}, including non-local operators 
constructed using up to two derivatives. This enables us to extract states confidently 
with spins up to $J=7/2$ for both even and odd parities. The two derivative operators also include 
operators that contains the field strength tensor appearing in it. A state having strong 
overlaps with these operators indicates the strong intrinsic gluonic content in it, and 
such a state is called a hybrid state~\cite{Edwards:2011jj}. Lattice operators are obtained by 
subducing these continuum operators on to various irreps of the symmetry of the lattice 
\cite{Edwards:2011jj}. For each of these irreps, we compute $N\times N$ matrices of 
correlation functions, where $N$ is the number of lattice operators used in each irrep. 
A subset of operators that are formed just by considering only the upper two components 
of the four component Dirac-spinors are called non-relativistic as they form the whole 
set of creation operators (with $SU(6)\otimes O(3)$ symmetry) in a leading order velocity expansion.

Lattice computations of hadron masses proceed through the calculations of the Euclidean two 
point correlation functions, between creation operators at time $t_i$ and annihilation operators at 
time $t_f$,
\vspace{-0.1cm}
\beq
C_{ij}(t_f-t_i) = \langle 0|O_j(t_f)\bar{O}_i(t_i)|0\rangle = \sum_{n} {Z^{n*}_iZ^n_j\over 2 m_n} e^{-m_n(t_f-t_i)} . 
\eeq{2pt} 

\vspace{-0.1cm}

\noindent The RHS is the spectral decomposition of such two point functions where the sum is over a discrete 
set of states. $Z^{n} = \langle 0|O_i^{\dagger}|n\rangle$ is the vacuum state matrix element, also called an
overlap factor. We employ a variational method \cite{Dudek:2007wv} to extract the spectrum of baryon states 
from the matrix of correlation functions constructed using a large basis of interpolating operators. 
The method proceeds by solving a generalized eigenvalue problem of the form 
\vspace{-0.1cm}  
\beq
\nonumber C_{ij}(t)v_j^{(n)}(t,t_0) = \lambda^{(n)}(t,t_0)C_{ij}(t_0)v_j^{(n)}(t,t_0),
\eeq{a10} 
where the eigenvalues, $\lambda^{(n)}(t,t_0)$ form the principal correlators and the eigenvectors are related 
to the overlap factors as $Z_i^{(n)} = \langle0|O_i|n\rangle = \sqrt{2E_n}\exp^{E_nt_0/2}v_j^{(n)\dagger}C_{ji}(t_0)$.
The energies are determined by fitting the principal correlators, while the spin identification of the states
are made by using these overlap factors as discussed in ref.\cite{Dudek:2007wv}.

\vspace*{-0.3cm}
\section{Results}
\vspace*{-0.2cm}
In Figure 1 we show the spin identified spectra of the triply charmed baryons where $3/2$ times 
the mass of $\eta_c$ is subtracted to account for the difference in the charm quark content \cite{Padmanath:2013zfa}. 
It is preferable to compare the energy splittings between the states, as it reduces the systematic uncertainty 
in the determination of the charm quark mass parameter in the lattice action and to lessen the effect of 
ambiguity in 
\begin{wrapfigure}{r}{80mm}
\vspace*{-0.2in}
\begin{center}
  \includegraphics[scale=0.3]{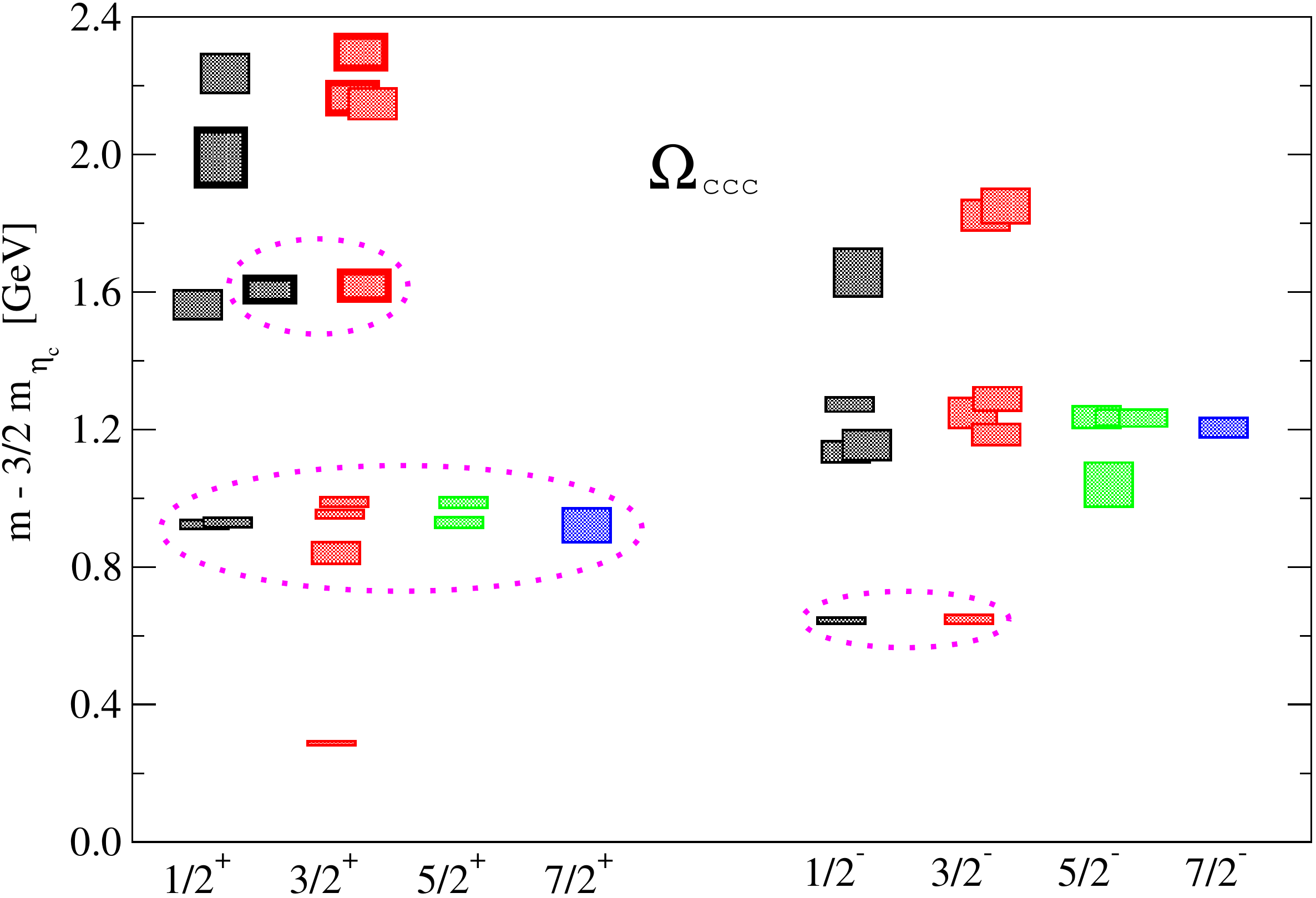}\\
\caption{Spin identified spectra of triply-charmed baryons with respect to $\frac32 m_{\eta_c}$.
The boxes with thick borders corresponds to the states with strong overlap with hybrid operators.
The states inside the pink ellipses are those with relatively large overlap to non-relativistic
operators.}
\end{center}
\vspace*{-0.2in}
\label{ccc_spectrum}
\end{wrapfigure}
the scale setting procedure. Boxes with thicker borders correspond to those states with a greater overlap 
onto the operators that are proportional to the field strength tensor, which might consequently be hybrid 
states. The states inside the pink ellipses have relatively large overlap with non-relativistic operators. 
A remarkable feature one can observe from Figure 1 is that though we use a large set of operators including many relativistic ones,
the number of low lying states in the non-relativistic bands exactly agree with expectations from models with an $SU(6)\otimes O(3)$ symmetry.

\fgn{ccq_spectrum} shows the spin identified spectra of the doubly charmed baryons \cite{Padmanath:2013laa}. Here the spectra
is shown with the mass of $\eta_c$ subtracted. The boxes and the pink ellipses represent similar quantities as in Figure 1.
Here again one can see the agreement between the number of states in the lower non-relativistic bands 
and the expectations as per a model with SU(6)$\otimes$O(3) symmetry.

\bef[bh]
\small
\parbox{.5\linewidth}{
\centering
  \includegraphics[scale=0.3]{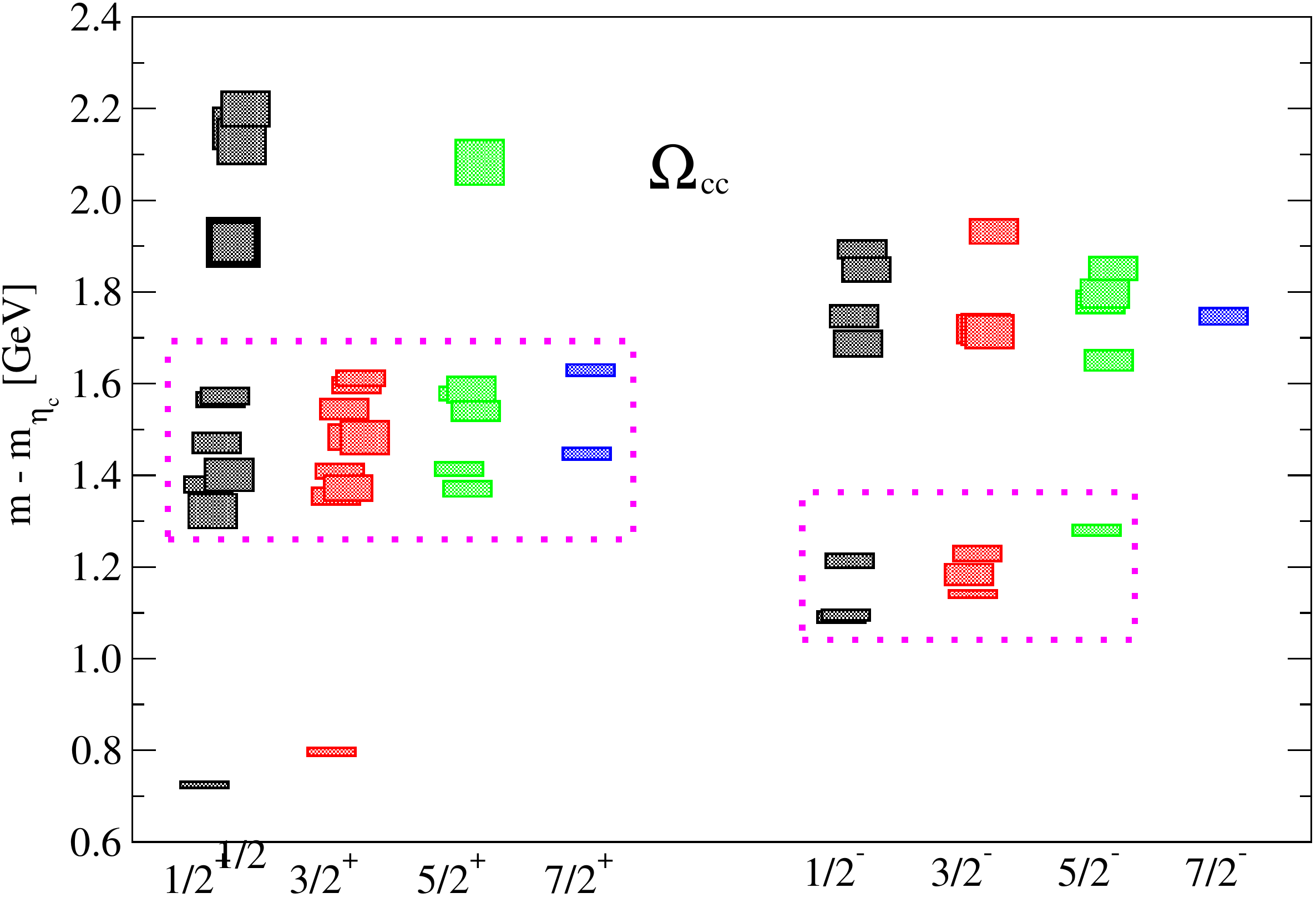}\\
(a)}
\parbox{.5\linewidth}{ 
\centering
  \includegraphics[scale=0.3]{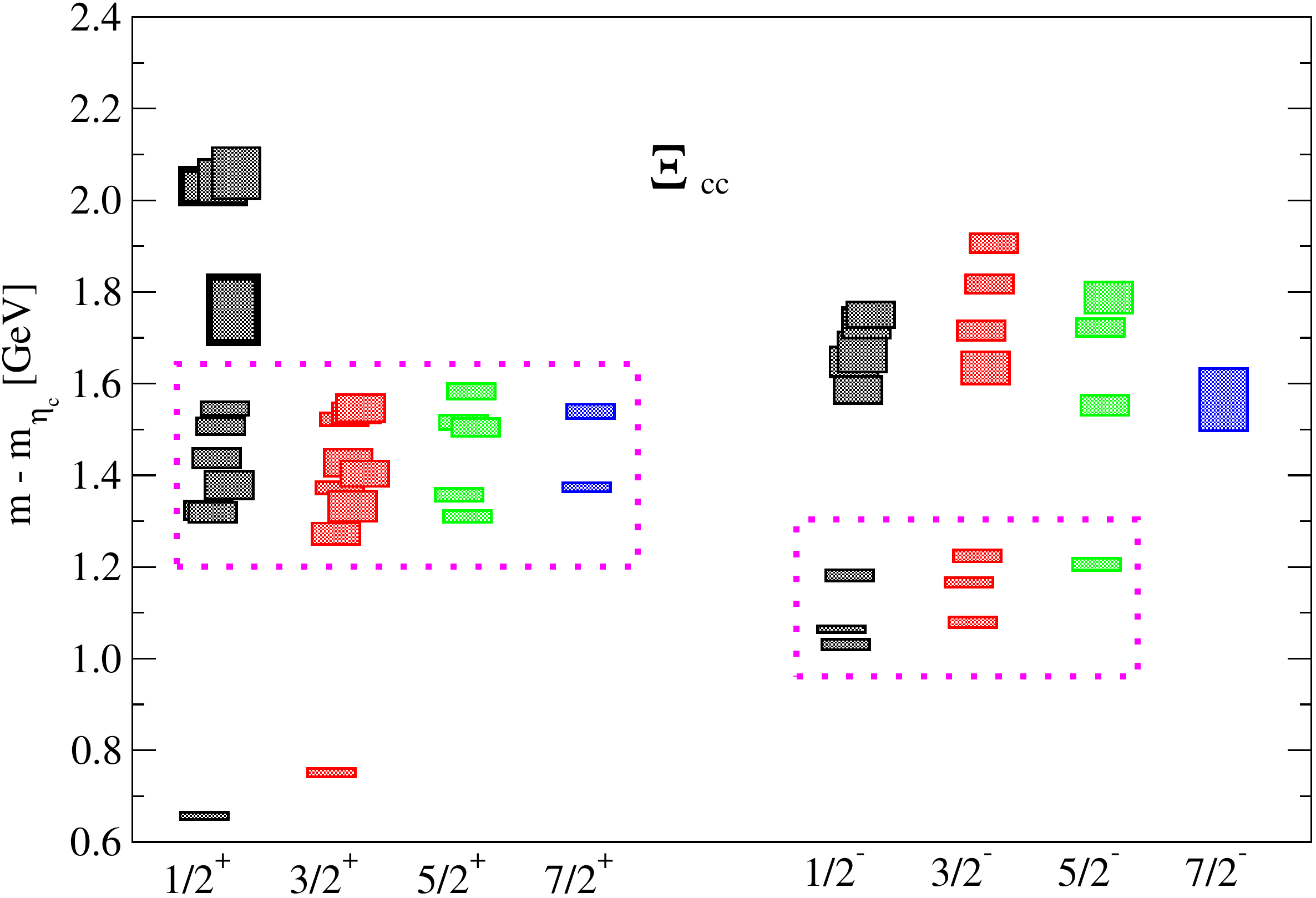}\\
(b)}
\caption{ Spin identified spectra of (a) $\Omega_{cc}$ and (b) $\Xi_{cc}$ baryon for both parities 
and with spins up to $\frac72$ {\it w.r.t.} $m_{\eta_c}$. The keys are same as in Figure 1.}
\eef{ccq_spectrum}

\bef[th]
\vspace*{-0.5cm}
\small
\parbox{.5\linewidth}{
\centering
  \includegraphics[scale=0.3]{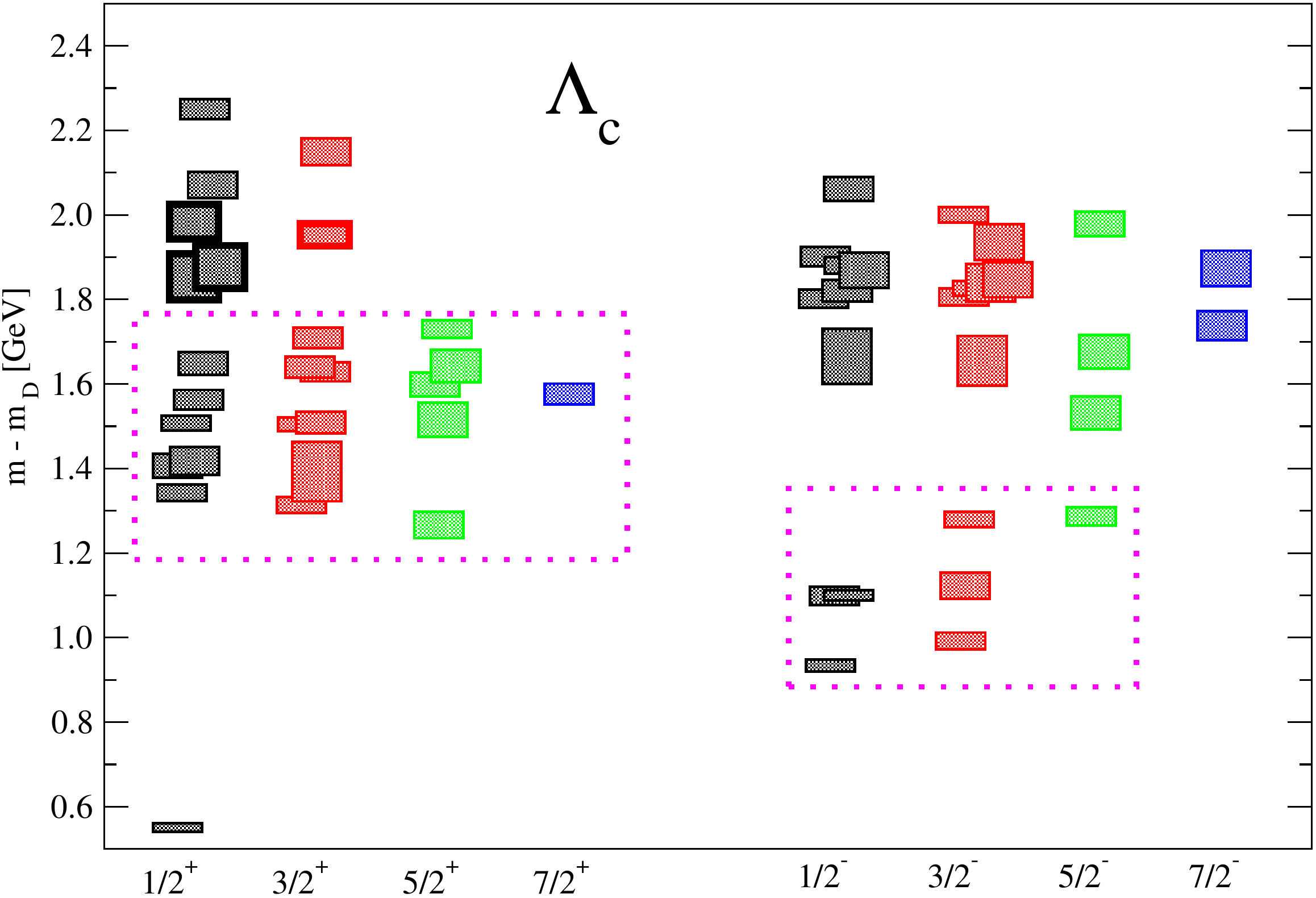}\\
(a)}
\parbox{.5\linewidth}{ 
\centering
  \includegraphics[scale=0.3]{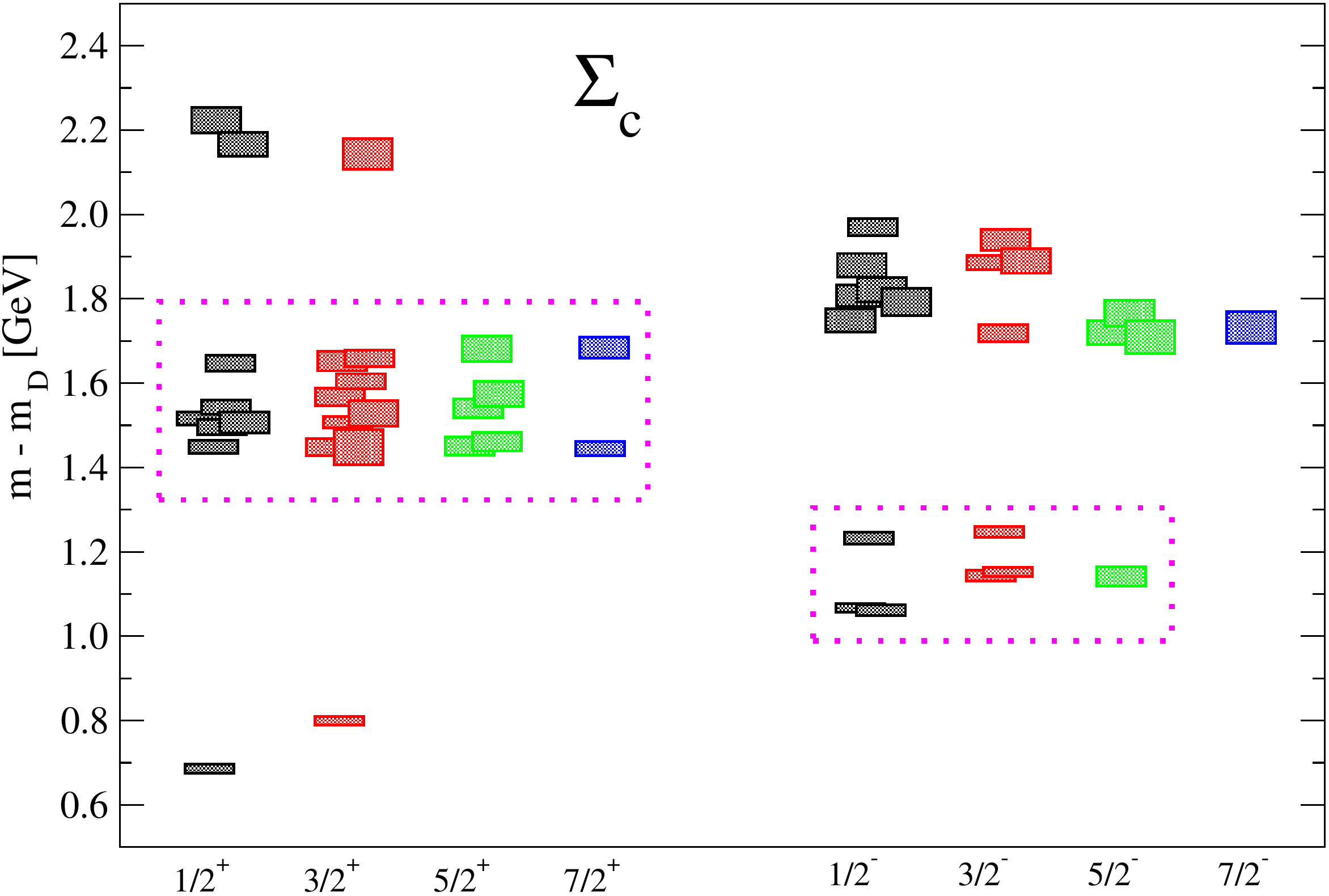}\\
(b)}
\parbox{.5\linewidth}{
\centering
  \includegraphics[scale=0.3]{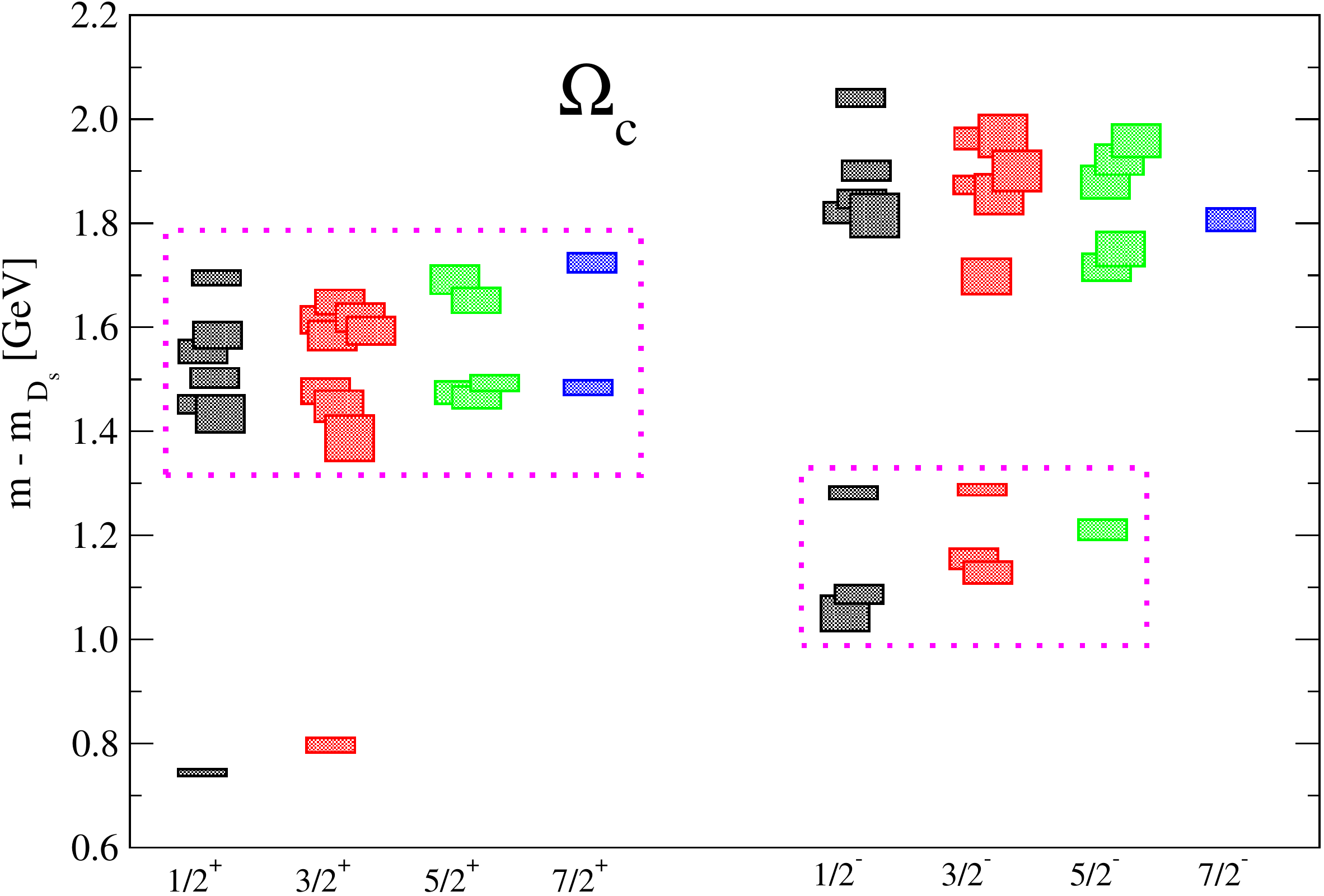}\\
(c)}
\parbox{.5\linewidth}{ 
\centering
  \includegraphics[scale=0.3]{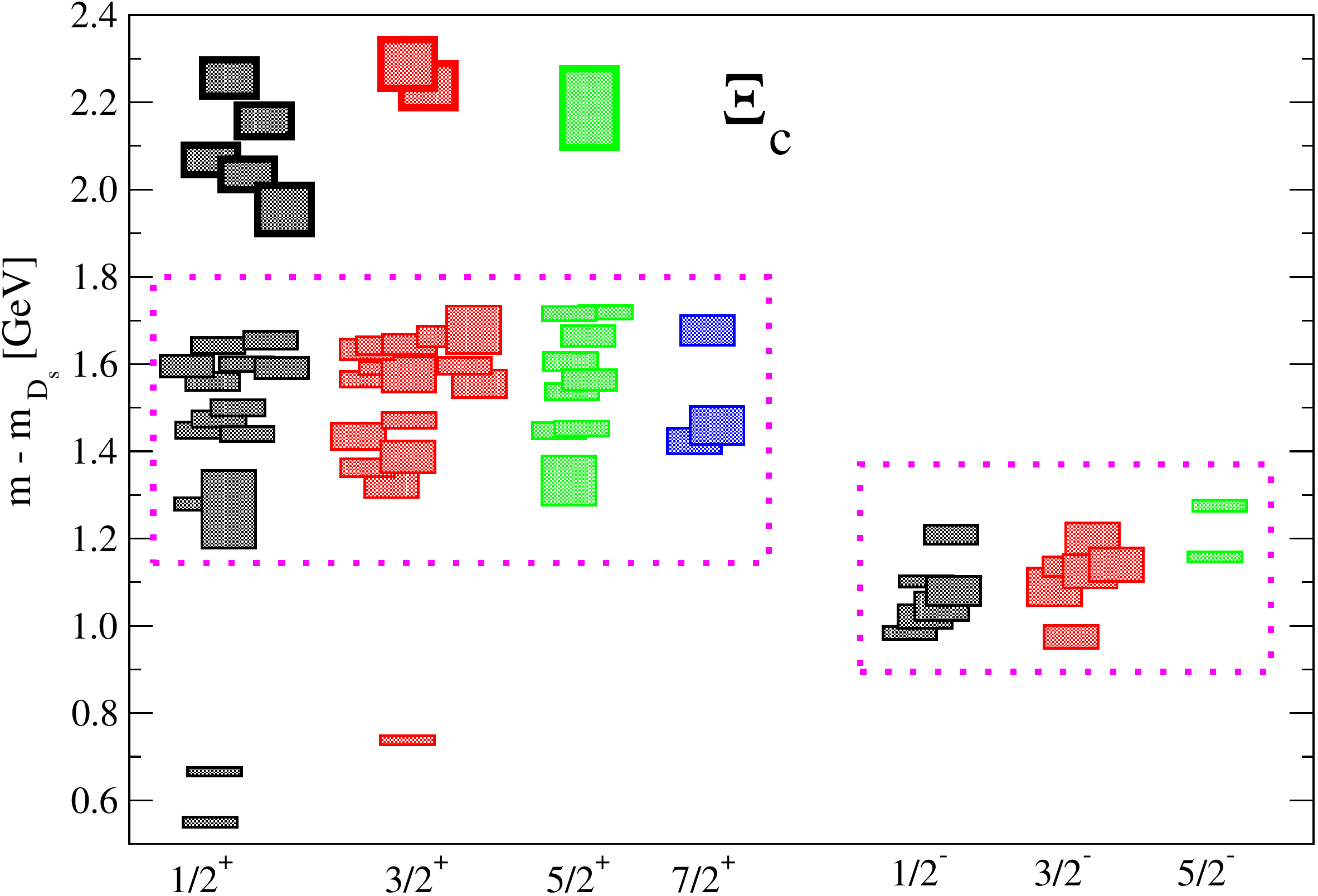}\\
(d)}
\caption{ Preliminary results on the spin identified spectra of (a) $\Lambda_{c}$, (b) $\Sigma_{c}$, (c) $\Omega_{c}$ and (d) $\Xi_{c}$ baryons for both parities {\it w.r.t.} $m_D$ (upper two) and $m_{D_s}$ (lower two) mesons. The keys are same as in Figure 1.}
\eef{LS_spectrum}

\fgn{LS_spectrum} shows the spin identified spectra of the singly charmed 
baryons, which include $\Lambda_c$, $\Sigma_c$, $\Xi_c$ and $\Omega_c$. Here for $\Lambda_c$ and
$\Sigma_c$ the spectra are shown with $m_D$ subtracted, while for $\Xi_c$ and $\Omega_c$ we plot 
the difference of the baryon mass from $m_{D_s}$ in order to account for the heavy flavor content. 
Here also there is good agreement between the number of states in the lower non-relativistic bands 
and the expectations as per a model with SU(6)$\otimes$O(3) symmetry.

\vspace*{-0.4cm}
\section{Conclusions}
In this work we present a comprehensive calculation on the ground and excited state spectra of 
singly, doubly and triply-charmed baryons by using dynamical lattice QCD. Preliminary results on 
singly charmed baryons are shown here for the first time. The spectra that we obtain have 
states with well-defined total spins up to 7/2 and the low lying states remarkably resemble the 
expectations of quantum numbers from $SU(6)\otimes O(3)$ symmetry. However, it is to be noted 
that we only mentioned statistical error in this work and the systematics from other sources 
like chiral extrapolation, lattice spacing are not addressed here. Also we have not incorporated 
multi-hadron operators which may effect some of the above conclusions, though to a lesser extent 
than their influence in the light hadron spectra. One other caveat in this work, particularly for 
singly charmed baryons, is that pseudoscalar mass used is unphysical ($m_{\pi}=$391 MeV).

\vspace*{-0.4cm}
\Acknowledgements
We thank our colleagues within the Hadron Spectrum Collaboration.  Chroma~
\cite{Edwards:2004sx} and QUDA~\cite{Clark:2009wm,Babich:2010mu} were used to 
perform this work on the Gaggle and Brood clusters of the Department of 
Theoretical Physics, Tata Institute of Fundamental Research and at Lonsdale 
cluster maintained by the Trinity Centre for High Performance Computing and at 
Jefferson Laboratory.

\end{document}